*2024-2025 CRA Quadrennial Paper*

# Now More Than Ever, Foundational AI Research and Infrastructure Depends on the Federal Government

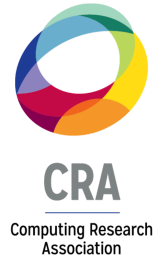
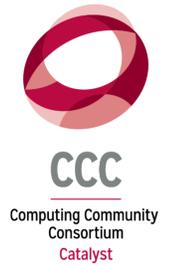

Michela Taufer (University of Tennessee, Knoxville), Rada Mihalcea (University of Michigan), Matthew Turk (Toyota Technological Institute at Chicago), Dan Lopresti (Lehigh University), Adam Wierman (California Institute of Technology), Kevin Butler (University of Florida), Sven Koenig (University of California, Irvine), David Danks (University of California San Diego), William Gropp (University of Illinois Urbana–Champaign), Manish Parashar (University of Utah), Yolanda Gil (University of Southern California), Bill Regli (University of Maryland), Rajmohan Rajaraman (Northeastern University), David Jensen (University of Massachusetts Amherst), Nadya Bliss (Arizona State University), Mary Lou Maher (Computing Research Association)

> **Leadership in the field of AI is vital for our nation's economy and security. Maintaining this leadership requires investments by the federal government. The federal investment in foundation AI research is essential for U.S. leadership in the field. Providing accessible AI infrastructure will benefit everyone. Now is the time to increase the federal support, which will be complementary to, and help drive, the nation's high-tech industry investments.**

There is enormous potential for artificial intelligence (AI) to benefit society. However, a technological gulf remains between the impressive, early-yet-business-centered achievements of AI and its enormous potential to address our pressing national problems, transform scientific discovery, enhance national security, improve industrial efficiency, and drive economic growth. Current AI tools are effective at summarizing information drawn from diverse sources but are also prone to generating so-called "hallucinations" (reasonable-looking but incorrect answers). They can be trained to solve complex mathematical problems in specialized domains, yet often fail at simple logical reasoning and struggle with open-ended challenges. Significantly, AI systems can reinforce the biases inherent in training data, leading to serious concerns about their reliability.

To build the next generation of AI that will be safe, smart, reliable, and energy-efficient, it will not be sufficient to simply increase model sizes or train on larger datasets. It is critical to make foundational advances leading to new AI architectures and paradigms (e.g., post-transformer models, long-range planning, continual learning, thinking beyond AGI,)[1] improve our understanding of the limitations and capabilities of AI, develop novel energy-efficient designs, and implement human-AI interfaces that ensure the safety, security, and reliability of AI systems. These advances will require a sustained commitment to investments in foundational AI research at America's key science agencies, in partnership with academia and industry, much like the decades of federally supported research in the field that date as far back as the 1950s.

Given that it is impossible to predict where the next great AI breakthrough will come from, it is necessary for the federal government to maintain a broad and diverse portfolio of investments in basic AI research. There is little reason to think that the marketplace alone will broaden the AI ecosystem to include ideas that would be hard to monetize outside the commercial sphere, including applications for national security and the public benefit.

AI technology is an exceptionally valuable tool to aid the federal government in meeting many national priorities. However, limits on federal investment in foundational AI research risk the U.S. falling behind our peer competitors like China, which is making massive investments in AI foundations and is already leading the world in the number of research papers that appear in top AI venues, especially in the subfields of machine learning, computer vision, and robotics. Referring to technical papers about language and reasoning models recently developed by Chinese startup DeepSeek, Chris Manning of Stanford University, a prominent figure in natural language processing, has stated that "we are in this bizarre world where the best way to learn about LLMs is to read papers by Chinese companies."

As part of this investment, the federal government must play a central, essential role in democratizing access to AI resources through infrastructure support for research at a range of scales. The federal government obviously cannot, and should not, attempt to compete with private industry. But only the government has the reach and resources to establish, fund, and coordinate comprehensive efforts that enable wider access to the resources necessary for cutting-edge AI research, design, development, and testing, and to leverage AI to address critical societal grand challenges. These governmental efforts should be focused on supporting the organizations and problems that cannot, or will not, be addressed by the large technology companies that increasingly dominate AI resources. Therefore, democratization of AI resources would create a powerful portfolio of efforts at different scales, including individual researchers, research groups, multi-institution centers, and small businesses, among other research stakeholders, which would foster a diversity of the research in the field's portfolio.

Small and medium-sized businesses require resources to build the next great AI product. Researchers require resources to pursue innovative AI hardware architectures, algorithms, and software. Civil society organizations need resources to ensure that AI benefits all, not just the few. Educational institutions, including community colleges, tribal colleges, and minority-serving institutions, require resources to ensure that the next generation of the workforce is fully AI-ready. State and local governments require resources to harness AI for public services and benefits. The federal government has the opportunity to empower large parts of American society to advance, innovate, and leverage cutting-edge AI ideas and technologies, but only if it acts to ensure appropriate resources for these AI innovators.

---

[1] For more information, see the AAAI 2025 Presidential Panel on the Future of AI Research report.

In practice, this effort will require sustained financial and intellectual investments over multiple years, just like the decades of support for fundamental research that has gotten the AI field to where it is today. As just one example, the [2025 Turing Award](#) (the Nobel prize in computing) was given to Andrew Barto and Richard Sutton for their foundational work, dating back to the 1990s, in reinforcement learning, a technology at the heart of all major AI systems today. Barto's research was supported through grants from the U.S. National Science Foundation (NSF) programs in robotics, robust intelligence, AI, and cognitive science, which have driven the long-term, fundamental advances in machine learning.

It is not enough to simply have a one-time purchase of chips and then move on to the next big idea. There needs to be commitments of ongoing funding to ensure state-of-the-art technologies are available to all; otherwise, the resource will run the risk of being plentiful but obsolete. There needs to be continuing funds for the development and delivery of educational and training materials, support staff, workshops, courses, and more about AI methods, data, and systems, or else user access to AI tools will be biased toward those with significant resources. There needs to be long-lasting support systems to ensure these resources are truly accessible to all.

## The Importance of Foundational AI Research

In order to maintain leadership in AI technology and its future critical applications, it is imperative to advance foundational research that aims to thoroughly understand current AI models, their capabilities, and their potential for good and for harm. What we do not know about current large AI models is a foundational understanding of:

- How and why these models scale and generalize

- How to know the amounts and kinds of training data needed to ensure accurate and robust results

- How to learn from limited data

- How to address faulty or bad data

- How to interpret and explain a model's decisions
- How to prevent adversarial attacks that result in harm
- How to align model behavior with American values and ethics.

These research topics need to be studied to ensure that the AI systems of the future are trustworthy, secure, risk-aware, energy-efficient, and human-centered.

These topics require the deep understanding that comes from fundamental, long-term research that is not focused on near-term applications or on fixing current AI models. Without increased emphasis on fundamental questions, progress in AI will eventually stall, allowing other countries that are increasing their investments in AI research to take the lead. Relying on U.S. industry to lead the way may bear fruit in the short run, but will eventually lose to those countries investing in long-term research. The nation risks falling behind in developing innovative, safe, and trustworthy AI and thus compromising our ability to address national and global challenges.

## The Importance of Democratized AI Resources

AI is being embraced across industries in the United States, driving a rapid increase in the resources required to advance the current limits of AI — including hardware, software, data, talent, and funding. These resource requirements pose a significant threat to U.S. innovation in AI if only a few organizations can pursue new ideas.

A few major technology companies drive innovation and research at an unprecedented scale in the AI ecosystem. For example, Nvidia is the world leader in AI chip manufacturing, controlling approximately 80 percent of the AI chip market. Likewise, the global cloud market is largely supported by a few key providers such as Amazon, Google, and Microsoft, which offer the computing and storage capabilities that enable cutting-edge AI applications worldwide, controlling over two-thirds of the $600 billion global cloud market. This overreliance on a small group of companies extends to the workforce, as top researchers are increasingly drawn to them. As just one data point, the percentage of AI PhDs taking jobs in industry (rather than academia) jumped from 41 percent in 2011 to 71 percent in 2022. This flight to industry highlights the challenge for researchers, including academic institutions, in recruiting and retaining talent, and risks stunting the training of the next generation of AI researchers.

Groups outside the giant technology companies — academic researchers or small businesses, state and local governments or non-profit organizations — too often lack the necessary financial, personnel, or computational resources to pursue innovative AI. For example, OpenAI's GPT-4 is reported to have cost over $100 million to develop, and hundreds of

thousands of dollars per day to run. Alternatively, DeepSeek has claimed that their system was trained much more cheaply; even if those claims are true, it still required substantial investment that are beyond the resources of most researchers. Few organizations can afford to compete in this space, even if they have access to the necessary computational resources. Other players are increasingly forced to tailor their hardware, software, and products to the needs of large technology companies, as they are the gateway to broad adoption. Given that we don't know where the next great AI innovations will come from, and the marketplace alone is ill-suited to broaden the AI ecosystem to include ideas that lack a direct financial incentive, the federal government's interests in national security and the public good make it well equipped to maintain a broad and diverse portfolio of investments in basic AI research.

Market dominance is not intrinsically problematic, but the current state of resources for AI research, design, and development raises significant risks for the American people. Most resources will inevitably be concentrated on the problems that matter to large technology companies, with datasets and models primarily tailored to their needs. Overall, the restricted focus on only a few problems means that innovation will be limited to those issues. Companies are often more incentivized to focus on market share, rather than important matters such as privacy or security. National competitiveness across every dimension will suffer if AI innovation is constrained in these ways.

Even worse, we might not even realize how AI innovation has been stunted precisely because of the concentration of resources. Efforts to independently evaluate or assess the AI models produced by large technology companies require access to the cutting-edge platforms on which those AI systems run. However, large technology companies have little incentive to provide resources to those who might compete with them. Limiting the ability to participate in the development and applications of AI can also lead to AI models and applications that are biased and lack fairness and representation.

## NAAIR and FASST

There are currently two nascent efforts to create some of the required AI R&D infrastructure. However, neither has been fully authorized or appropriated: the National AI Research Resource (NAIRR), and the Department of Energy's Frontiers in Artificial Intelligence for Science, Security, and Technology (FASST) program. NAIRR is designed to provide AI resources broadly to researchers and innovators across sectors and problems. At the same time, FASST aims to support deep research efforts specifically on DOE-relevant scientific and national security challenges.

These efforts complement one another, and both should be fully funded without delay. Both programs would help to democratize access to AI research and development resources,

2024-2025 CRA Quadrennial Paper    5

though in different ways. Both represent paths for the federal government to accelerate innovation and insights in AI, much as the government has historically done in astronomy, physics, and other fields.

These two efforts are complementary to the NSF-funded AI Institutes program and serve to democratize resources for AI innovation. Extending the AI Institutes program along with additional public-private partnerships should be explored for particular sectors or challenges, supported by relevant agencies and departments. Such efforts could be:

- AI Grand Challenges: Develop public participation mechanisms to solicit ideas for an ambitious AI Grand Challenges agenda. It could be updated regularly to reflect new priorities and opportunities within the subfields of AI.

- Develop a national cyberinfrastructure that extends beyond metropolitan areas, reaching rural communities to empower local populations, ensure last-mile sustainability, and enable all Americans to engage with and benefit from AI.

- Develop services to support the creation of AI-ready data, democratize its discovery, integration, and use, and access its quality and impacts.

- Develop a "marketplace" to connect data and problem owners (including the federal government) with students across a range of educational institutions for capstone and thesis projects.

- Provide financial incentives for infrastructure providers to provide training and compute to regional, under-resourced organizations, including small businesses.

- Develop software libraries to facilitate the creation of trustworthy AI systems in small- and medium-sized organizations.

- Promote industry-academia partnerships to build a sustainable, long-term workforce pipeline that equips individuals with practical AI knowledge to meet national industry needs.

We find ourselves at a critical juncture in the development of AI. We can continue going down the path of resource consolidation in the hands of a few companies. Or we can choose to support the democratization of AI, which will unleash the creative and innovative potential of American enterprise. NAIRR and FASST are critical first steps, but also only starting points given the magnitude of the impact and the attendant disruptions we will face during the coming AI revolution. We must continue to explore and develop additional mechanisms to dramatically expand the people, communities, and organizations who have access to the resources required to truly innovate. This latter path is the only way to ensure that we have trustworthy, innovative

AI that benefits the American public, improves national and economic security, and helps us maintain our position as global leaders in innovation.

## Summary and Recommendations

To maintain U.S. leadership and competitiveness in AI, it is essential to both 1) secure increased and sustained national funding for fundamental AI research, and 2) provide a commonly available set of AI research tools and resources for all to use. We define fundamental AI broadly, encompassing not only the technical aspects of model development but also the foundational research necessary for advancing the socio-technical (or the interaction between people, technology, and the environment) dimensions of AI. These two efforts are complementary and will build off each other.

Below, we outline key areas of focus for fundamental AI research and infrastructure.

- Develop new architectures for AI that integrate multiple models of computational intelligence to overcome the inefficiencies, lack of trust, and lack of explainability in current large language models. This requires sustained funding for high-risk approaches that lead to long-term solutions.

- Pursue alternatives to the predominant models in machine learning that leverage our understanding of intelligent systems in all forms. This requires funding for highly exploratory and interdisciplinary research.

- Extend the basic scope of AI to include collaborative intelligence among multiple intelligent systems and humans. This requires supporting a broader portfolio of research, going beyond research on human-AI interaction to include researchers from cognitive, brain, and computational scientists.

- Many of the challenges of building effective AI systems are not purely technical, but involve difficulties in the interactions between humans, both individually and collectively, and AI-enabled technologies. These socio-technical challenges require different research approaches to bridge the human and the technical involving additional disciplines like social science and economics that are not well represented in our current AI funding models.

- AI education and workforce issues are critical, and the U.S. needs more AI professionals. Sustained funding for AI education and workforce development is necessary to retrain existing workers, reskill displaced workers, and develop systems for improving AI education.

- The federal government should incentivize a baseline of common educational and training materials that all users could utilize. This would make cross-sector training and AI education simpler, as it would limit companies from monopolizing essential training tools or resources.

- NAIRR and FASST need to have full, bipartisan support and should be fully funded. Additionally, public-private partnerships should be explored for particular national challenges, supported by relevant agencies and departments through national efforts.

---

*This quadrennial paper is part of a series compiled every four years by the **Computing Research Association (CRA)** and members of the computing research community to inform policymakers, community members, and the public about key research opportunities in areas of national priority. The selected topics reflect mutual interests across various subdisciplines within the computing research field. These papers explore potential research directions, challenges, and recommendations. The opinions expressed are those of the authors and CRA and do not represent the views of the organizations with which they are affiliated.*

*This material is based upon work supported by the U.S. National Science Foundation (NSF) under Grant No. 2300842. Any opinions, findings, and conclusions or recommendations expressed in this material are those of the authors and do not necessarily reflect the views of NSF.*